\documentclass[letterpaper]{article} 

\usepackage{aaai24}  

\usepackage{times}  
\usepackage{helvet}  
\usepackage{courier}  
\usepackage[hyphens]{url}  
\usepackage{graphicx} 
\def\UrlFont{\rm}  
\usepackage{natbib}  
\usepackage{caption} 
\usepackage{algorithm}
\usepackage{algorithmic}

\usepackage[bottom]{footmisc}

\usepackage{newfloat}
\usepackage{listings}
\DeclareCaptionStyle{ruled}{labelfont=normalfont,labelsep=colon,strut=off} 
\floatstyle{ruled}
\newfloat{listing}{tb}{lst}{}
\floatname{listing}{Listing}
\usepackage{graphicx} 
\usepackage{array}    
\usepackage{makecell} 
\usepackage{booktabs} 
\usepackage[table,xcdraw]{xcolor}   

\usepackage{multirow}
\usepackage{arydshln}

\usepackage{hyperref}  
\nocopyright 

\title{Societal Adaptation to Advanced AI}

\author{
    Jamie Bernardi\textsuperscript{\rm 1}\thanks{Equal contribution. Order of first three authors randomised. Authors are free to list themselves first author in their CVs.},
    Gabriel Mukobi\textsuperscript{\rm 2 *},
    Hilary Greaves\textsuperscript{\rm 3 *},
    Lennart Heim\textsuperscript{\rm 1},
    Markus Anderljung\textsuperscript{\rm 1 *}\thanks{Senior author.}
}
\affiliations{
    \textsuperscript{\rm 1}Centre for the Governance of AI\ \ 
    \textsuperscript{\rm 2}Stanford University\ \ 
    \textsuperscript{\rm 3}University of Oxford\ \  

    contact@jamiebernardi.com, gmukobi@cs.stanford.edu, hilary.greaves@philosophy.ox.ac.uk, \\lennart.heim@governance.ai, markus.anderljung@governance.ai
}

\begin{document}
\pagenumbering{arabic}
\maketitle

\begin{abstract}
Existing strategies for managing risks from advanced AI systems often focus on affecting what AI systems are developed and how they diffuse. However, this approach becomes less feasible as the number of developers of advanced AI grows, and impedes beneficial use-cases as well as harmful ones. In response, we urge a complementary approach: increasing societal adaptation to advanced AI, that is, reducing the expected negative impacts from a given level of diffusion of a given AI capability. We introduce a conceptual framework which helps identify adaptive interventions that avoid, defend against and remedy potentially harmful uses of AI systems, illustrated with examples in election manipulation, cyberterrorism, and loss of control to AI decision-makers. We discuss a three-step cycle that society can implement to adapt to AI. Increasing society’s ability to implement this cycle builds its resilience to advanced AI. We conclude with concrete recommendations for governments, industry, and third-parties.
\end{abstract}

\section{Introduction}
\label{id:h.wpo743gvspwr}





The diffusion of advanced AI—AI systems that approach and exceed human capabilities—brings both benefits and risks, necessitating careful governance. Many existing approaches to managing AI risks focus on identifying potentially harmful capabilities of AI systems~\citep{shevlane2023modelevaluation} and modifying how those capabilities are developed and made available. Examples include monitoring inputs and outputs to block harmful prompts and responses~\citep{openai2023gpt4}, regulating deployment~\citep{anderljung2023frontier}, or employing training methods to generate safer outputs~\citep{bai2022training}. We refer to interventions of these types as \textit{capability-modifying interventions}.

As the cost of developing advanced AI decreases, however, it becomes less feasible for risk management to rely solely on capability-modifying interventions~\citep{pilz2023increased}. Government oversight of ever-smaller actors’ development and deployment activities would be both difficult and undesirable~\citep{sastry2024computingpower}. Moreover, capability-modifying interventions already fail to comprehensively address risk, and in addition restrict beneficial as well as harmful applications.

Capability-modifying interventions should therefore be complemented by \textit{adaptation} to advanced AI: adjusting other aspects of society to reduce the expected negative impacts \textit{downstream} of capability diffusion, holding fixed which AI capabilities are created and how they diffuse.\footnote{An analogous distinction between capability modification and adaptation is already well-recognised in efforts to address climate change. Climate mitigation, like capability modification, tackles risk at its source: here, reducing net CO\textsubscript{2} emissions in order to prevent consequent climate change. Climate adaptation is a matter of adjusting society to reduce the impact of climate change that nonetheless does occur~\citep{ipcc2014glossary}. The~\citep{OECD2023ClimateFinance} estimates that 27\% of total climate finance is spent on adaptation.

In general, adaptation (to advanced AI, and generally) includes seizing opportunities to increase benefits gained, as well as avoiding downsides. However, in this paper we will focus on adaptation to avoid downsides.}

While a large portion of the efforts aimed at addressing the risks of AI systems with relatively modest capabilities focus on adaptation measures, efforts to address the risks from \textit{more advanced }AI systems tend to predominantly focus on capability-modifying interventions. We urge an increased focus on adaptation to advanced AI as a crucial complement to capability-modification.

This paper motivates the need for societal adaptation to advanced AI (Section~\ref{sec:need}) and introduces a framework for conceptualising such adaptation (Section~\ref{sec:framework}). We apply this framework to three examples of AI risk (Section~\ref{sec:examples}). We explore the structures society needs to successfully adapt, introducing \textit{resilience} as the \textit{capacity to adapt} (Section~\ref{sec:resilience}). Section~\ref{sec:rec} offers recommendations for government, industry, academia, and nonprofits.

\section{The Need for Societal AI Adaptation}
\label{sec:need}

New technologies often introduce novel risks. These risks can arise from the intentional misuse of the technology, or as an unintended consequence. Over time, society typically adapts to the risks. This trajectory can be observed in the historical rise and fall of pedestrian road collisions~\citep{DfT2022PedestrianFactsheet} and numbers of smokers~\citep{ritchie2013smoking} in the United Kingdom, for example.

Though adaptation does to some extent arise spontaneously, it usually benefits from deliberate planning and effort. Pedestrian fatalities have decreased in part due to speed limits~\citep{jepson2022speedlimits} and road safety campaigns\footnote{https://www.think.gov.uk/. Accessed 2024-05-13.}, not only increased pedestrian caution. Adaptation can be reactive, responding to harm as it manifests, or proactive by anticipating potential risks~\citep{ipcc2001glossary}.

In this section, we motivate the need for adaptation to complement capability-modifying approaches to risks from advanced AI. We suggest that capability-modifying approaches will become less feasible and less effective over time (Section~\ref{sec:need:less-effect}), and we explain how adaptation may also aid beneficial diffusion of advanced AI (Section~\ref{sec:need:benefits}).

\subsection{Capability-Modifying Approaches Will Become Less Effective Over Time}
\label{sec:need:less-effect}

\subsubsection{Increased Diffusion Makes Capability-Modifying Interventions Less Feasible}
\label{sec:need:less-feasible}

\citet{pilz2023increased} observe that the cost of training an AI system to a given level of performance has been decreasing over the last decade, due to efficiency improvements in training algorithms and in hardware performance, and that these trends are likely to continue. In 2020, training OpenAI’s GPT-3 was estimated to cost at least \$4.6 million in computing costs~\citep{li2020gpt3}; two years later, Mosaic claimed to offer the same performance for a tenth of the cost~\citep{venigalla2022gpt3quality}. \citet{rahman2024tracking} estimate that 56 models have now been trained using more compute than GPT-3, by 29 organisations. In sum, we should expect that over time, more actors will have the resources required to train advanced and potentially risky AI systems.

Increased access to developing advanced AI technologies enables significant benefits (Section~\ref{sec:need:open}). But beyond a certain point, it undermines the feasibility of AI governance approaches that solely rely on capability-modification~\citep{scharre2024futureproofing}. If capability-focused interventions focused on an absolute level of capability, they would affect a growing number of small actors, someday potentially including individual citizens. This would be both impractical and undesirable. Capability-modifying approaches focused on relative rather than absolute performance may remain more feasible. Nonetheless, such approaches would have to be accompanied by adaptive measures.

\subsubsection{Safeguards Are Not Failsafe}
\label{sec:need:safeguards}

Models are often deployed with capability-modifying safeguards, such as fine-tuning~\citep{bai2022training}  or input and output filtering~\citep{openai2023gpt4}. But \textit{solely} relying on such safeguards is insufficient for managing risks, for the following reasons.

\textbf{Some proportion of developers will deploy models without safeguards}, e.g. because such safeguards can affect product quality. For instance, Mistral releases some of its model weights without safeguards to “empower users to test and refine moderation\footnote{http://docs.mistral.ai/getting-started/open\_weight\_models. Accessed: 2024-05-13.}.” As the number of actors developing models increases (Section~\ref{sec:need:less-feasible}), so too will the diversity of decisions developers make regarding safeguards. Even if some countries mandate safeguards, other, more permissive, regimes will likely remain.

\textbf{Some safeguards can be cheaply removed by small-scale actors.} Even if models are initially deployed with safeguards, it can be cheap for small teams with access to model weights to intentionally remove safeguards~\citep{yang2023shadowalignment, gade2023badllama}. Additionally, even without access to weights, techniques like jailbreaking, as in~\citet{anil2024many}, can circumvent many existing safeguards.\footnote{There are, however, other safeguards that are more difficult to undo, such as unlearning~\citep{li2024wmdp} or model fingerprints that could aid traceability~\citep{lukas2019fingerprinting}}

\textbf{Model leakage and theft.} Even if model weights are secured to prevent safeguard tampering, models (and their dangerous capabilities) could still be leaked or stolen via information security failures~\citep{nevo2023modelweights}.\footnote{Whilst weights were not \textit{stolen} in this case, Meta’s Llama was leaked online one week after it was made available to researchers on-request~\citep{vincent2023metaai}.}

\begin{figure*}[t!]
  \centering
  \includegraphics[width=1.000\linewidth]{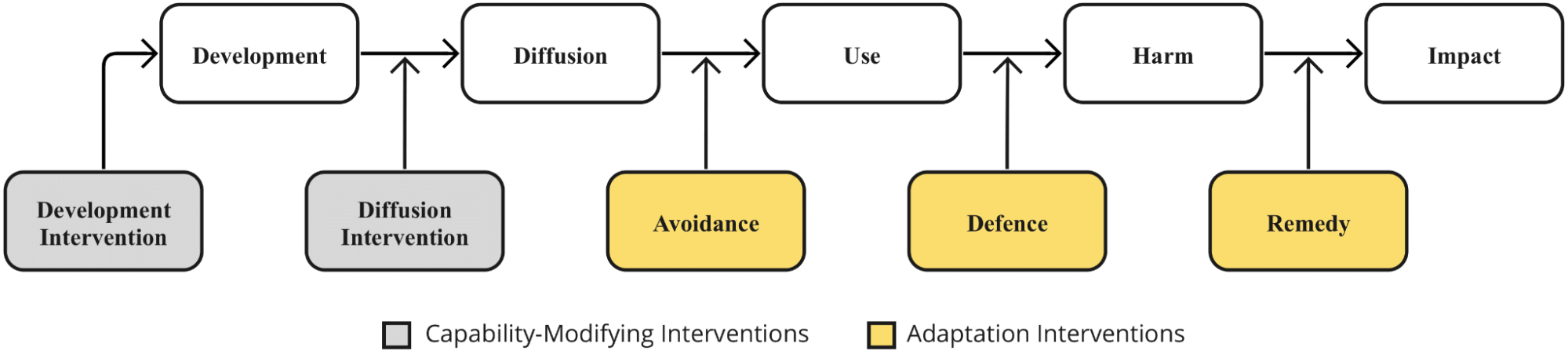}
  \caption{A simplified causal pathway to an AI system causing negative impacts and how various types of intervention can reduce them. The focus of this paper is on the latter three interventions: adaptation interventions.}
  \label{fig:pathway}
\end{figure*}

While AI safeguard failures appear to have relatively limited impacts today, we should be prepared for greater potential impact in the case of future, more advanced systems. Such preparations will require complementing capability-modifying interventions with adaptive ones.

\subsection{Adaptive Approaches Aid Beneficial Diffusion}
\label{sec:need:benefits}

Societal adaptation to advanced AI may be not only necessary, but also beneficial in other ways, through promoting the diffusion of AI capabilities and the open development of AI systems.

\subsubsection{Adaptation Can Enable Beneficial Use}
\label{id:h.qyboplfudkkd}

While capability-modifying interventions can reduce risk, they will often be blunt instruments, since they inhibit beneficial use-cases as well as harmful ones, resulting in a Use-Misuse Tradeoff~\citep{anderljung2023protecting, weidinger2023sociotechnical}. For example, restricting an AI system’s knowledge of virology through techniques like unlearning~\citep{li2024wmdp} or filtering-out API requests and model responses related to that capability~\citep{openai2023gpt4} could reduce the hypothesised risk of enabling bioterrorists~\citep{nelson2023risks}, but may also hinder students’ and scientists’ ability to learn and to combat diseases. To the extent that society is able to adapt, we would be better positioned to harness the benefits from such dual-use capabilities without incurring unacceptable risks.

\subsubsection{Adaptation Can Enable Open Development}
\label{sec:need:open}

Open development of AI systems, particularly the open release of model weights, can be both beneficial and harmful~\citep{seger2023opensourcing, kapoor2024impact}. Benefits include stimulating innovation~\citep{langenkamp2022opensource}, distributing decision-making power, mitigating market concentration~\citep{vipra2023marketconcentration, ukcma2023ai}, and facilitating external scrutiny of models~\citep{bucknall2023access}. On the other hand, open-weight models limit safeguarding options (Section~\ref{sec:need:safeguards}) and have caused tangible harms already, such as the production of DeepFakes depicting non-consensual intimate imagery~\citep{lakatos2023revealing} and AI generated Child Sexual Abuse Material (CSAM)~\citep{iwf2023abuse}.

Without societal adaptation, the primary approaches for avoiding unacceptable levels of harm from open deployment involve restricting openness. An adaptive approach offers more promise of realising the benefits of openness while simultaneously reducing its harms.

\section{A Framework for AI Adaptation}
\label{sec:framework}

In the previous section, we argued that addressing the risks from advanced AI is not only a matter of intervening on AI capabilities, but also ensuring society’s \textit{adaptation}: reducing the expected negative impacts from advanced AI, holding fixed which AI capabilities are developed and how they diffuse.

In this section, we offer a framework to guide thinking about such adaptation. The framework lays out the structure of a causal chain leading to negative impacts from AI,\footnote{A \textit{threat model }is a model of a particular possible causal pathway. Section~\ref{sec:framework} lays out the abstract structure; Section~\ref{sec:examples} discusses three example threat models.} and offers a categorisation of interventions that could reduce such impacts.\footnote{The “use,” “initial harm,” “impact” distinction we use is similar, but not identical, to distinctions often used in legal scholarship between a “wrong” (an inappropriate action taken by some party), “injury” (a harmful event), and “damage” (the magnitude of impact of an injury)~\citep{nolan2013negligence}, and in risk management between “cause”, “event” and “consequence”~\citep{waycott2018managing}. In reality, in any given case there are a huge number of causal steps leading to harm, which could be mapped onto this framework in various equally valid ways. \label{fn7}}

\subsection{The Causal Chain to Negative Impacts of AI}
\label{sec:framework:causal-chain}

Negative impacts from AI systems follow the causal pathway illustrated in Figure~\ref{fig:pathway}: 

\textbf{Development:} An AI capability or system is developed. 

\textbf{Diffusion:} The capability or system becomes available to various users.

\textbf{Use:}\footnote{A more general term would be “operation” of the AI system: in some cases of loss of control over AI, there need not be a “user.”} The AI system is used in a way that could cause harm. This harm could be actively intended (“misuse”), such as a cybercriminal using a new general-purpose model to automate the generation of spear phishing messages, aiming to access sensitive information on a company’s systems. It could also be that AI is used in a way that has a concerning likelihood of causing unintentional harm (“accident”).

\textbf{Initial harm:} The use of the AI system results in some proximate harmful event.\footnote{More precisely, we might define “initial harm” as roughly an “event that would,\textit{ by default}, leave some party worse off or have some right of theirs violated”. The clause “by default” leaves open that remedial action might prevent the party in question from \textit{actually }experiencing negative impacts at the end of the day. It is also consistent with our usage that “harm” occurs in cases in which that harm causally leads to more than adequate compensation, so that the \textit{net }eventual effect is positive.} In the case of misuse, this can be thought of as the \textit{initial success }of the malign use of AI: success in the first step of the actor’s plan.\footnote{In an accident case, what counts as the “initial harm” in a given case is (still) more open to stipulation (cf. footnote\ref{fn7}).} (In our example, the “initial harm” occurs if the cybercriminal \textit{succeeds} in gaining access to the sensitive information in question.)

\textbf{Impact: }The initial harm results in further negative impact. This impact could be measured in terms of e.g. lives lost, economic opportunities lost, or damage to national security.\footnote{In the examples we’ll consider, the impact will most often be negative, but it could be made zero or even positive given sufficiently effective adaptation. For example, people losing their jobs due to AI could receive financial compensation that exceeded their employment income, thereby making them (at least financially) better off.} (In our example, the cybercriminal might sell sensitive information from an arms manufacturer’s systems to a state actor that either reproduces a weapon or learns how to exploit its weaknesses, thereby leading to additional lives lost.)

To apply the framework in practice, it is often best to fix a specific use, harm, or impact of interest, and then identify the other steps accordingly. For example, if focussing on the harm of unauthorised access to sensitive computer systems, one might consider a range of uses that may lead to such breaches (e.g. spear phishing or insider threats), and a range of impacts such access might have (e.g. stealing important data or harming citizens in a cyberattack on physical infrastructure).

\subsection{Interventions to Reduce Negative Impacts}
\label{id:h.lq03iv76ur8}

To reduce negative impacts, policymakers can intervene at different points along this causal chain.

\subsubsection{Capability-Modifying Interventions}
\label{id:h.6h1eukr93s1m}

\textit{Capability-modifying} interventions intervene at points immediately preceding the “development” and “diffusion” steps:

\textbf{Development interventions.} Society can affect which AI capabilities are developed. For example, companies could refrain from developing systems that have certain potentially harmful capabilities, or make systems that are more resistant to jailbreaking, have higher chances of refusing potentially harmful requests, or have outputs that can be more easily identifiable as AI-generated.

\textbf{Diffusion interventions.} Society can affect which AI systems are made available, to whom, and with what degrees of access. For example, companies can employ “staged release”: gradually making the system more widely available~\citep{solaiman2023gradient}. They could make potentially risky models available only via an API, allowing them to implement secure safeguards, such as watermarking or content provenance tags~\citep{shevlane2022structuredaccess}. They could enforce terms of service policies, removing access from customers who use the system in prohibited ways.

\subsubsection{Adaptation Interventions}
\label{id:h.sxz3xi947bh3}

\textit{Adaptation interventions, }the primary focus of this paper, intervene at later stages in the causal chain. Such interventions immediately precede the “use”, “initial harm” or “impact” stages of that chain. (Occasionally, a specific intervention can affect multiple points along the causal chain.)

\textbf{Avoidance interventions.} Society can reduce the expected extent of the potentially harmful use of AI, making the problematic actions in question more difficult to engage in, or more costly compared to relevant alternatives.\footnote{These two routes to avoidance correspond to the distinction between “deterrence by denial” and “deterrence by punishment” that is commonly drawn in the literature on military strategy~\citep{mazarr2018deterrence}.} One can make it more difficult for a given instance of potentially harmful AI activity to occur by limiting the user’s or the AI system’s access to key resources that are required for the activity in question, or to key actuators that are required for completion of the intended action. (In the spear phishing example we introduced in Section~\ref{sec:framework:causal-chain}: relevant companies could make it harder for cybercriminals to access the names and contact details of their staff.) One can make potentially harmful uses of AI more \textit{ex ante} costly by building institutions that create credible threats of punishment for harmful use.\footnote{The main means of creating such a credible threat is of course to actually issue after-the-fact penalties. Superficially, this makes it look as though systems of penalty act later in the causal chain; but their crucial disincentive effect acts at the “avoidance” point.}



\begin{table*}[t!]
    \centering
    \renewcommand{\arraystretch}{1.3}
    \begin{tabular}{m{3cm}p{5.5cm}p{6.5cm}}
        \toprule
        \multicolumn{1}{c}{\textbf{Risk}} &
        \multicolumn{1}{c}{\textbf{Threat Model}} &
        \multicolumn{1}{c}{\textbf{Example Adaptations}} \\
        \midrule

        \multirow{3}{=}{
            \parbox{1\linewidth}{
                \vspace{-5pt}
                \centering
                \includegraphics[width=1.cm]{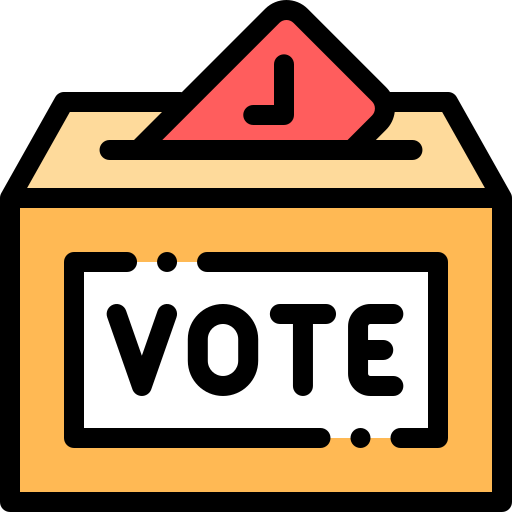} \\
                \textbf{Election Manipulation with Generative AI (\ref{sec:examples:election})}
            }
        }
        & \textbf{Use:} AI misuse to create synthetic election manipulation media. 
        & \textbf{Avoidance:} Criminalising election interference, verifying humanity on social media. \\
        \cdashline{2-3}[1.5pt/3pt] 
        & \textbf{Harm:} Voters misled, holding false election beliefs. 
        & \textbf{Defence:} Public awareness campaigns, content provenance, AI content detection tools. \\
        \cdashline{2-3}[1.5pt/3pt] 
        & \textbf{Impact:} Disenfranchisement, misrepresentation, political instability. 
        & \textbf{Remedy:} Transparent investigations into electoral integrity, rerunning elections in extreme. \\

        \midrule

        \multirow{3}{=}{
            \parbox{1\linewidth}{
                \vspace{-5pt}
                \centering
                \includegraphics[width=1.cm]{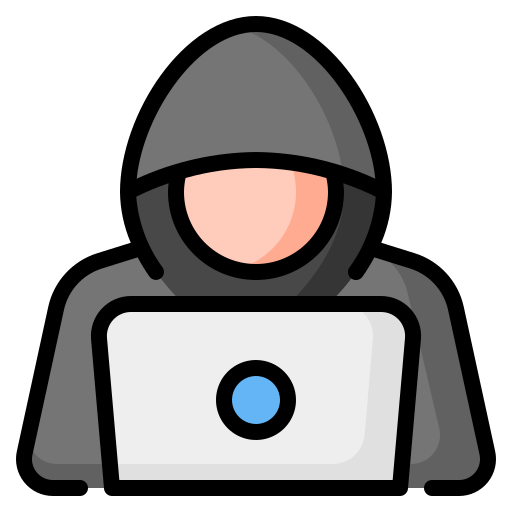} \\
                \textbf{Cyberterrorism Attacks on Critical Infrastructure (\ref{sec:examples:cyber})}
            }
        }
        & \textbf{Use:} AI aids non-state actors in cyberattacking critical infrastructure. 
        & \textbf{Avoidance:} International justice agreements, enhanced detection of cyber intrusions. \\
        \cdashline{2-3}[1.5pt/3pt] 
        & \textbf{Harm:} Critical infrastructure taken offline or damaged, data theft. 
        & \textbf{Defence:} AI-enhanced cyber defence, information-sharing networks. \\
        \cdashline{2-3}[1.5pt/3pt] 
        & \textbf{Impact:} Loss of life, economic damage, national security threats. 
        & \textbf{Remedy:} Compensation schemes, redundancy in critical infrastructure, rapid repair plans. \\

        \midrule

        \multirow{3}{=}{
            \parbox{1\linewidth}{
                \vspace{-5pt}
                \centering
                \includegraphics[width=1.cm]{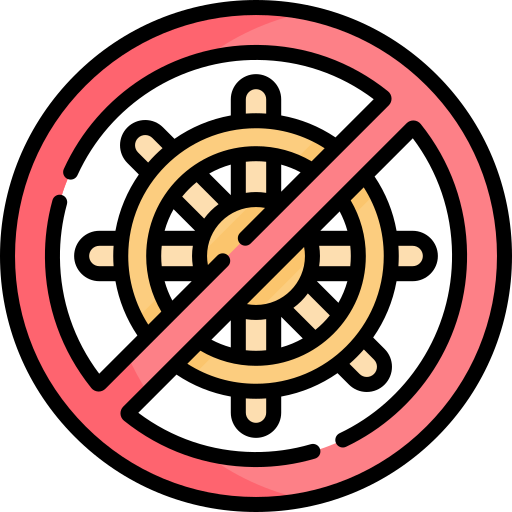} \\
                \textbf{Loss of Control to AI Decision- Makers (\ref{sec:examples:loc})}
            }
        }
        & \textbf{Use:} Increased reliance on AI in general decision-making.
        & \textbf{Avoidance:} Regulating automation in high-stakes decision-making. \\
        \cdashline{2-3}[1.5pt/3pt] 
        & \textbf{Harm:} High-stakes decisions made without effective human oversight. 
        & \textbf{Defence:} Human-in-the-loop requirements, rigorous auditing, whistleblower protections. \\
        \cdashline{2-3}[1.5pt/3pt] 
        & \textbf{Impact:} Harmful decisions, potentially catastrophic loss of societal control. 
        & \textbf{Remedy:} Disempowering harmful AI decision-makers, shared incident reporting. \\

        \bottomrule
    \end{tabular}
    \caption{Examples of adapting to AI risks. Each is described in more depth in Section~\ref{sec:examples}. Images: Flaticon.com}
    \label{tab:examples}
\end{table*}

\textbf{Defence interventions.} Holding fixed that the potentially harmful use of AI occurs, society can reduce the expected extent of the corresponding initial harm. In our spear phishing example, “defence” is a matter of reducing the chance that the spear phishing emails succeed in giving the cybercriminal access to the sensitive information. For example, companies could provide anti-phishing training to their staff, and implement tools to warn staff of suspected phishing emails. They could ensure that only a very small number of staff members have access to particularly sensitive information, and then only with approval from other employees.

\textbf{Remedial interventions.} Holding fixed that the initial harm occurs, society can reduce or eliminate the expected negative impact downstream of that. In our spear phishing example, this might go via reducing the extent to which national security is undermined as a result of the sale of the proprietary information to a foreign actor. For example, the company could include some false and misleading documents on its servers. Governments could reduce incentives for staff with relevant implicit knowledge to work for the foreign actor, on the grounds that implicit knowledge is often required to complement information contained in documents.

\section{Examples of Adapting to AI Risks}
\label{sec:examples}

To illustrate the practical application of the framework described in Section~\ref{sec:framework}, we discuss three examples of AI threats and corresponding adaptations shown in Table~\ref{tab:examples}: election manipulation, cybersecurity, and gradual loss of control to AI decision-makers. For each example, we describe a concrete threat model for how harm might occur and list some possible adaptation interventions categorised by our framework of \textit{Avoidance}, \textit{Defence}, and \textit{Remedy}. We do not make claims about the likelihood or importance of these AI threats, or the merits of the particular adaptive interventions we suggest; the goal is rather to illustrate how the framework presented in the previous section might aid brainstorming of possible adaptive interventions.

\subsection{Election Manipulation with Generative AI}
\label{sec:examples:election}

\subsubsection{Threat Model}
\label{id:h.drxydpih0zxq}

Generative AI systems can create high-quality text~\citep{jones2024turing}, video~\citep{brooks2024video}, and audio media.\footnote{https://elevenlabs.io. Accessed: 2024-05-08.} This synthetic media is often difficult to distinguish from authentic content~\citep{cooke2024detection}, and frontier language models are already about as persuasive as humans~\citep{goldstein2024propaganda, durmus2024persuasiveness}. Further, the capabilities for generating high-quality synthetic media are already quite diffuse, including access to proprietary\footnote{https://openai.com/index/gpt-4o-and-more-tools-to-chatgpt-free. Accessed: 2024-05-13.} as well as open access LLMs\footnote{https://llama.meta.com/llama3. Accessed: 2024-05-13.}, and image generators.\footnote{https://www.midjourney.com. Accessed: 2024-05-13. https://stability.ai/news/stable-diffusion-3. Accessed: 2024-05-13.}

\textbf{Use:} Generative AI systems might be used maliciously to manipulate democratic elections. For example, synthetic media could impersonate political figures for defamatory political content~\citep{meyer2023deepfake}. This disinformation could be micro-targeted to individual voters for greater efficacy~\citep{salvi2024persuasiveness}, especially if people increasingly use and trust personalised AI companions~\citep{roose2024aifriends}.

\textbf{Initial Harm:} We take the initial harm to be: voters holding false election-relevant views that they otherwise would not hold. For example, they could believe that the impersonations are genuine and that the fake news stories are true~\citep{west2023ai2024elections}, believe incorrect information about when and where they can vote~\citep{swenson2024bidenrobocall}, or be manipulated into weighing the merits of candidates in a different way than their authentic selves would.

\textbf{Impact:} A misled and manipulated electorate negatively impacts the validity and efficacy of democratic elections, diverging an election’s outcome from the values and will of the voters. AI-enabled election manipulation could also undermine public trust in elections, which can in turn can lead to political instability. Furthermore, a society where it is widely believed that AI-generated and authentic content are indistinguishable is vulnerable to the “liar’s dividend,” where public figures may dismiss real incriminating evidence as fake~\citep{chesney2019deepfakes, schiff2023liars}.

\subsubsection{Adaptation Examples}
\label{id:h.4bsmaolnolo9}

\textbf{Avoidance:} Governments can deter election interference by criminalising it~\citep{lerner2023misinformation}, subject to requirements of free speech~\citep{toney2024aclu}. Social media platforms can require some "proof of humanity" for creation of user accounts, making it more challenging for bot accounts to spread disinformation~\citep{shoemaker2024humanity}.

\textbf{Defence:} Public awareness campaigns can empower individuals to critically assess AI-generated content. Content provenance techniques throughout the lifetime of a piece of media, e.g. when a photo is captured and edited, can help to verify genuine content~\citep{srinivasan2024fingerprints, earnshaw2023fighting}. AI content detection tools can enable platforms to take appropriate actions such as removal, labelling, or adding scalable counter-disinformation such as Community Notes~\citep{wojcik2022birdwatch}.

\textbf{Remedy:} In extreme circumstances, given robust evidence of election manipulation, governments could rerun elections, as has been done in Germany~\citep{martin2024berlin}, India~\citep{agarwala2024rerun}, Malawi~\citep{kell2020malawi} and Serbia~\citep{gec2024serbia}, though caution is required~\citep{huefner2007remedying}. Impartial and transparent investigations into the integrity of the electoral process can build public trust to avoid secondary harms from a disgruntled public.

\subsection{AI-Enabled Cyberterrorism Attacks on Critical Infrastructure}
\label{sec:examples:cyber}

\subsubsection{Threat Model}
\label{id:h.nmwh07ush3c}

Increasingly capable large language models could lower the barriers to cyberattacks by rapidly finding and exploiting vulnerabilities~\citep{li2024wmdp, fang2024llm}; though see also~\citep{rohlf2024llm}.

\textbf{Use: }Future advanced AI systems could aid small non-state actors, such as terrorist groups, to carry out cyberattacks on critical infrastructure necessary for societal security, safety, and stability~\citep{newman2024cybersecurity}. These non-state actors may be more willing to carry out such attacks, because of having low accountability and/or fear of retaliation compared to nation-states.

\textbf{Initial Harm: }Such attacks could take critical infrastructure offline or cause lasting damage to it. State-level cyberattacks unaided by AI have already been used to disable electrical grids in Ukrainian cities~\citep{finkle2016ukraine} and undermine nuclear infrastructure in Iran~\citep{kushner2013arduino}. Cyberattackers targeting digital infrastructure have been used to steal large sums of money and exfiltrate sensitive information from governments~\citep{CSIS2024CyberIncidents}.

\textbf{Impact:} Critical infrastructure, by definition, is vital to societal needs. Damage to systems such as healthcare, energy, or communications could lead to enormous loss of life, economic damage, national security threats, or provocations toward international conflict.

\subsubsection{Adaptation Examples}
\label{id:h.xeucwleamrkg}

\textbf{Avoidance:} Robust international agreements against cyberterrorism could facilitate global cooperation in detecting, tracking, and prosecuting cyberterrorists~\citep{peters2020cybercrime}. Enhancing state abilities to detect cyber intrusions with access to critical infrastructure systems could preemptively identify and neutralise threats~\citep{cisa2013cyber}, especially including advanced persistent threats~\citep{cisa2024nation}.

\textbf{Defence:} Defensive AI capabilities can augment traditional cyber defence, for example by detecting and patching security vulnerabilities~\citep{lohn2022cyber}. Better information-sharing networks enhance the ability to detect diffuse or stealthy cyberterrorism and rapidly mitigate its impacts~\citep{johnson2016cyber}.

\textbf{Remedy:} Appropriate compensation schemes can reduce harm by spreading the costs associated with cyberattacks. Decoupled and redundant critical infrastructure, such as backup power for hospitals~\citep{davoudi2015power}, can ensure continuity of service. Cities can prepare to rapidly restore attacked infrastructure—for example, via planning and drills for rebooting the power grid or repairing compromised digital systems.

\subsection{Loss of Control to AI Decision-Makers}
\label{sec:examples:loc}

\subsubsection{Threat Model}
\label{id:h.9st1kj8iz3tv}

AI developers are increasingly building highly capable general-purpose AI systems that can carry out tasks without human supervision. OpenAI’s Charter explicitly commits to attempting the development of artificial general intelligence (AGI), defined as “highly autonomous systems that outperform humans at most economically valuable work”~\citep{openai2018charter}. As these systems increase in capability and see more widespread use, eventually there is a risk of “value erosion” and losing control of society to AI decision-makers~\citep{assadi2023humanity, dafoe2018governance}.

\textbf{Use:} Unlike misuse cases, the widespread use of AI decision-makers may arise without any individual intending harm. If AI systems seem more efficient or effective than human decision-makers, simple cost-benefit analyses may pressure institutions to rely more on AI~\citep{hendrycks2023selection}. For example, AI decision-makers could at some point replace board members in corporations~\citep{pugh2019ai}, policymakers in governments~\citep{samuel2019ai}, and commanders in militaries~\citep{clarke2022survey}. Furthermore, this reliance on AI decision-makers could compound: increasingly capable AI systems may produce work outputs and audit trails that are increasingly difficult for humans alone to supervise, leading to reliance on AI auditors~\citep{christiano2021alignment}.

\textbf{Initial Harm: }High-stakes decisions come to be made by AI alone on a large scale, without humans either in the decision loop or in a position to effectively oversee decisions.

\textbf{Impact: }While AI decision-makers could certainly bring many benefits~\citep{koster2022mechanism}, they could also cause harm by sometimes making much worse decisions than would be made by humans, even if they \textit{seem }better on average. Simple machine learning predictors may already exhibit algorithmic bias in high-stakes applications such as criminal justice~\citep{angwin2016bias}. Military decisions made by AI could escalate international conflicts~\citep{rivera2024escalation} or could lead to high rates of civilian casualties - as alleged by~\citep{abraham2024lavender}, especially if “automation bias” causes humans to defer more to AI~\citep{cummings2012automation}. Beyond bad decisions, overreliance on AI decision-makers could also lead to human enfeeblement~\citep{arvai2024hidden}. Ultimately, if AI decision-makers are misaligned to human-compatible goals, loss of control to AI could constitute an existential catastrophe~\citep{hendrycks2023risks, carlsmith2022power}.

\subsubsection{Adaptation Examples}
\label{id:h.mtx9zazesl3t}

\textbf{Avoidance:} Regulation could limit decision-making automation in certain high-stakes industries or government roles until these systems have been proved trustworthy~\citep{coy2024jobs}, similar to the existing regime of requiring trials for new pharmaceuticals~\citep{usfda2017drugreview}.

\textbf{Defence}\textbf{:} Human-in-the-loop requirements can require human oversight for certain high-stakes decisions, such as was proposed in the U.S. Block Nuclear Launch by Autonomous Artificial Intelligence Act of 2023~\citep{ussenate2023blocknuclear}, or ensure that AI decision-makers are augmenting and not strictly replacing humans~\citep{acemoglu2019automation}. Society could rigorously audit high-stakes provisional AI decisions before acting on them, and red team these auditing mechanisms. Whistleblower protections could encourage people to report issues in AI decision-making~\citep{katyal2018private, bloch-wehba2024whistleblowing}. Lastly, society could invest considerable resources in ensuring that AI systems do in fact act in accordance with our wishes, even where humans are incapable of providing effective supervision~\citep{bowman2022progress}.

\textbf{Remedy:} Government agencies could “bust” harmful AI decision-makers in critical roles, such as corporate executives, disempowering them similar to the way in which antitrust agencies bust corporate decisions that undermine consumer welfare. Shared incident reporting mechanisms could help institutions piece together diffuse patterns of failure~\citep{mcgregor2021cataloging}.

\section{Resilience: The Capacity to Adapt}
\label{sec:resilience}

\begin{figure}[h!]
  \centering
  \includegraphics[width=0.75\linewidth]{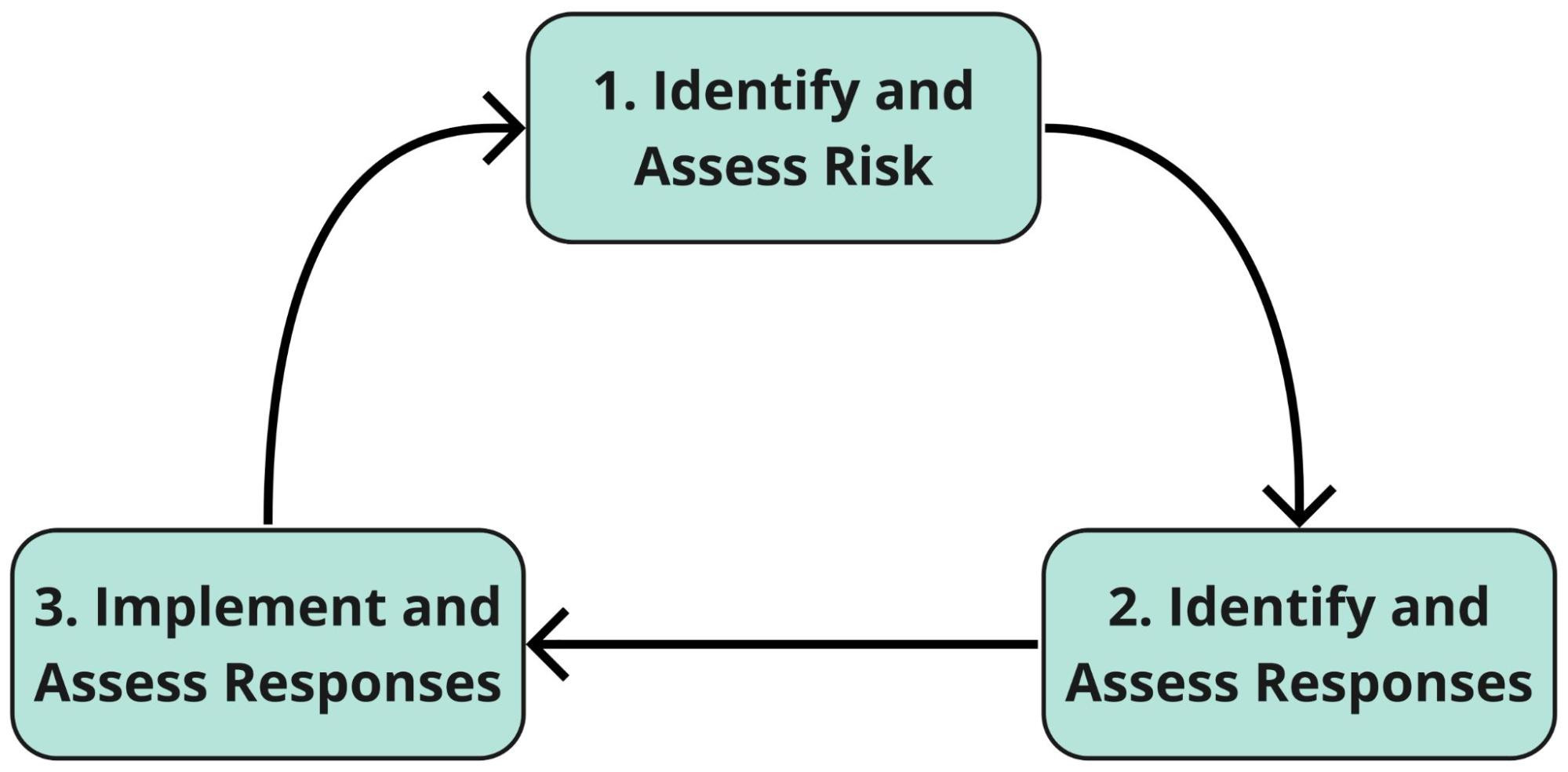}
  \caption{The three-step adaptation cycle that must be implemented to successfully adapt to advanced AI. Resilience is society’s capacity to perform this loop.}
  \label{fig:cycle}
\end{figure}

To ensure society adapts to advanced AI, certain structures and processes are required—specifically, continual implementation of the three-step cycle shown in Figure~\ref{fig:cycle}:

\begin{enumerate}
    \item Identify, forecast, and assess risks introduced or exacerbated by advanced AI systems.
    \item Identify and assess possible adaptive responses to address those risks.
    \item Implement appropriate adaptive responses and measure their effectiveness.
\end{enumerate}

Similar cycles describe the adaptation process in other contexts, for example climate change~\citep{eea2024adaptation}. However, the challenge of implementing this cycle effectively is especially acute in the case of advanced AI. This is due to the pace of technological change, the potential scale of impacts, and (in some cases) the indirectness of causal pathways from AI use to negative impacts.

We will say that a society with a strong capacity to adapt effectively is \textit{resilient} to advanced AI.\footnote{“Capacity to adapt” is one standard meaning of the term “resilience,” though there are others.} In this section, we describe each component of the adaptive cycle and outline possible initiatives for building society’s capacity to execute it.

\subsection{Identify, Forecast, and Assess Risks}
\label{id:h.lgiy1cbt1obr}

The planning of appropriate adaptations begins with a threat model: a mapping of the particular causal pathway by which a given AI system might lead to negative impacts. Such threat models should take into account the interests and views of all relevant stakeholders~\citep{watkins2021governing, lazar2023safety}.

Early availability of information helps make threat models more accurate, and provides relevant actors with more time to identify and implement adaptive responses. For example, in anticipation of the diffusion of AI-enabled vulnerability detection capabilities, DARPA and ARPA-H invested in hardening infrastructure against cyberattacks.\footnote{https://aicyberchallenge.com/.Accessed 2024-05-08.}

Adaptation-relevant information can be gathered at various points along the causal pathway to negative impacts from an AI system~\citep{stein2024rolegovernmentsincreasinginterconnected}:

\textbf{Pre-development information.} Before an AI system is developed, some information can be gathered to predict the type and extent of likely capabilities~\citep{kolt2024responsible, toner2023frontier}. Reporting on compute usage~\citep{sevilla2023compute, heim2024cloud} and registration of large training runs~\citep{hadfield2023registry, whitehouse2023executiveorder} can indicate in advance where novel capabilities are most likely to arise. Documenting datasets can help to predict unwanted model behaviours such as bias~\citep{gebru2018datasheets}.

\textbf{Pre-deployment information.} Before an AI system is deployed, labs and external parties can produce and share relevant information~\citep{anderljung2023aspire, mitchell2019modelcards}, e.g. by evaluating models for dangerous capabilities~\citep{shevlane2023modelevaluation} and examining human interactions with the system~\citep{weidinger2023sociotechnical}. AI developers can publish safety cases that assess whether deploying the system would impose unacceptable risks~\citep{clymer2024safety}.

\textbf{Integration and usage information.} After deployment, information can be gathered on where and how AI systems are being integrated, which may help society to predict where and how harmful use is most likely to occur~\citep{javadi2021monitoring, bonney2024ai} and understand its societal impact. Staged release protocols~\cite{solaiman2023gradient} could provide opportunities for monitoring use under limited release. Companies deploying advanced AI could be required to report their own aggregate usage statistics~\citep{kolt2024responsible}, and application providers could implement identifiers, real-time monitoring, and activity logging for AI agents~\cite{chan2024visibility}. Experiments can be set up to collect information on usage in plausibly representative samples~\citep{zhao2024wildchat}.

\textbf{Incident information.} This helps us recognise initial harms and negative impacts as they occur\footnote{https://oecd.ai/en/incidents. Accessed: 2024-05-08.}). One challenge in detecting harm from AI systems is that the causal roles of AI are often quite indirect, diffuse and unpredictable: far more so than for climate change or smoking, where the causal mechanisms are simpler and better understood. Watermarking or AI content identification tools~\citep{fernandez2023signature, ukdsit2023ai} could help to track where advanced AI systems have been used.

\subsection{Identify and Evaluate Possible Adaptive Responses}
\label{id:h.u1zaxp5qb98b}

Once a given threat model is sufficiently well-evidenced, society must identify plausible adaptive responses, and evaluate these to make an informed choice of which to implement.

To identify plausible responses, society can invest in research to identify ways in which a given threat model might effectively be blocked (via “avoidance”, “defence” or “remedy”, in terms of the framework we offered in Section~\ref{sec:framework}).\footnote{Our discussion in Section~\ref{sec:examples} illustrates in outline what this might look like for three examples, but for a proper treatment, vastly more detail and careful analysis is required.} Sometimes, this might require identifying possible new technologies, not yet developed, whose availability would enhance adaptation (consider the invention of airbags in response to the risk of car crashes). To evaluate proposed adaptive interventions, researchers must take into account cost-effectiveness, and impact on beneficial activity. As in the case of identifying possible pathways to harm, there are theoretical approaches to those calculations (e.g. modelling) and empirical approaches (e.g. controlled trials or natural experiments).

\subsection{Implement Adaptations and Measure Effectiveness}
\label{id:h.pq4ny8eaca5b}

Even when some groups in society are well-informed about risks and appropriate adaptive responses, adaptive interventions may not be put into practice, for at least the three reasons below. It took decades for society to implement measures to reduce smoking after its negative consequences were widely understood.

\textbf{Shared awareness and understanding. }Successful implementation requires shared awareness and understanding of appropriate adaptive interventions across society, including at least government, the private sector, academia and non-profit organisations, as well as (often) the general public. Effective communication between these sectors is therefore vital for ensuring identified adaptive responses are integrated into planning.

\textbf{Institutional capacity.} Successful implementation also depends on the existence of appropriate institutions for resolving collective action problems, and on organisations’ technical, financial and institutional capacities for monitoring and responding to AI risk. The rapid pace of development makes adaptation particularly challenging in the case of advanced AI, potentially raising challenges faster than society is equipped to implement solutions. This could be especially problematic if some risks are path-dependent, threatening permanent and irreversible damage when a pathway to harm proceeds unchecked, even temporarily. For example, a major disruption to the labour market could lead to mass dissatisfaction, economic difficulty, and resulting societal instability that would be harder to address than adapting to the initial stages of labour automation~\citep{klinova2021shared}.

\textbf{International coordination.} Appropriate international institutions and coordination may be required for effective collective action. To illustrate, difficulty in coordinating with AI Safety Institutes (which aim to evaluate frontier AI systems) in multiple jurisdictions has been cited as one underlying reason the UK AI Safety Institute has not received frontier model access prior to deployment~\citep{manancourt2024ai}. This has resulted in frontier model release without any public body conducting pre-deployment evaluation.

Once new adaptations have been implemented, their effectiveness should be monitored to assess whether the intervention should be scaled, changed, or dropped.\footnote{Conceptually, this is a return to the first step in the adaptation process: identify, forecast, and assess (remaining) societal risks from AI systems.}

\section{Recommendations}
\label{sec:rec}

To ensure society identifies, prioritises, and implements adaptations to AI, we highlight the following nine recommendations for decision-makers across policy, industry, academics and non-profits.

\subsection{Understanding-Based Recommendations}
\label{id:h.qs4njalu31bz}

\begin{itemize}
    \item \textbf{Measure and Predict AI Risks:} Governments should fund academics and auditors that measure and predict AI capabilities and corresponding risks, and build frameworks to ensure robust oversight of frontier AI companies~\citep{ee2023cybersecurity}. Frontier AI companies should carry out pre-deployment evaluations in collaboration with governments and third parties, reporting both development plans and deployment risks. They should make considerable investments to improve best practice in risk identification and mitigation. Academics should work to improve the science of risk and capabilities assessment. 
    \item \textbf{Build an External Scrutiny Ecosystem}\textbf{:} High-stakes AI development and deployment decisions should be informed by third-party assessments. Policymakers have an important role in ensuring such access is granted and that such external scrutiny is both informative and in fact informs important decisions~\citep{raji2022outsider, anderljung2023aspire}
    \item \textbf{Establish Incident Reporting Mechanisms:} Governments should establish incident reporting systems and requirements~\citep{walker2024merging}, along with whistleblower protections. Non-profit organisations can implement pilots of such programs~\citep{mcgregor2021cataloging}.
\end{itemize}

\subsection{Implementation Recommendations}
\label{id:h.ru38bf4lxre}
\begin{itemize}
    \item \textbf{Employ staged release: }AI companies should employ staged release protocols for their frontier systems, thereby giving society more time to implement adaptations~\citep{shevlane2022structuredaccess, solaiman2023gradient, seger2023opensourcing}.
    \item \textbf{Improve AI Literacy}\textbf{:} Educators, governments and journalists should continually make the general public, industry leaders, and key decision-makers aware of what advanced AI systems are capable of and their corresponding impacts~\citep{long2020ailiteracy}.
    \item \textbf{Sanction Known Harmful Uses:} Governments may need to criminalise certain harmful uses of advanced AI systems.
\end{itemize}

\subsection{Strategic Recommendations}
\label{id:h.ywu6j427y3tk}

\begin{itemize}
    \item \textbf{Use Defensive AI:} Governments should incentivize AI companies to develop and provide access to AI systems to defend against AI-caused threats~\citep{buterin2023optimism}. Such efforts may be bolstered by the fact that widely available AI systems may lag behind the capability of frontier systems~\citep{pilz2023increased}, which could differentially be put to defensive uses. For example, in cybersecurity, frontier systems that identify and fix vulnerabilities faster than widely diffused systems can exploit them could improve the offence-defence balance~\citep{lohn2022cyber, AIxCC2024}.
    \item \textbf{Secure International Cooperation}\textbf{:} Governments should facilitate international cooperation to increase adaptation~\citep{ho2023institutions}. For example, the various AI Safety Institutes and potentially the EU AI Office could coordinate to conduct pre-deployment testing of frontier AI systems to identify emerging risks and share information, thereby providing states with the time and knowledge to better adapt.
    \item \textbf{Invest in Adaptation}\textbf{: }Governments, philanthropists and private entities should allocate sufficient funds for timely societal adaptation to advanced AI. This could take many forms, such as funds for third-party organisations to build resilience~\citep{microsoft2024resilience}, funds to existing institutions to execute adaptation, or funds to establish new institutions focused on adaptation-specific needs such as red-teaming society for AI vulnerabilities or applying defensive AI.
\end{itemize}

\section{Conclusion}
\label{id:h.yktb0ig5xk3o}

As increasingly advanced AI systems are developed and widely diffused, society will need not only capability-modifying interventions, but also adaptation interventions to manage the accompanying risks. This is because capability-modifying interventions (i) become less feasible over time as it becomes possible for smaller and smaller actors to train advanced AI systems, (ii) are not failsafe, and (iii) inhibit beneficial as well as harmful uses.

This paper presents a framework for conceptualising and identifying the range of possible adaptive interventions in response to a given threat. Avoidance interventions allow that the AI capability in question has diffused, but inhibit dangerous uses of that capability. Defence interventions block or reduce the severity of initial harms along the pathway to negative impact, after dangerous use takes place. Remedial interventions intervene causally downstream of the initial harm, to diminish total negative impacts. We illustrated how this framework might aid brainstorming by applying it to three examples: election manipulation, cyberterrorism, and loss of control.

While some adaptation will happen by default, we expect sufficient adaptation will require deliberate action, foresight, and considerable investment. To adapt effectively, society will need to continually (i) identify and assess risks, (ii) identify and evaluate possible adaptations, and (iii) implement and measure the effectiveness of selected adaptations. We should increase society’s resilience to advanced AI, by increasing its capacity to execute this cycle.

\appendix
\section{Related Work}
\label{id:h.dtex8yb558ln}

In this paper we have urged the importance of:

\begin{itemize}
    \item Implementing \textit{adaptation }to advanced AI, defined as reducing the expected negative impacts from advanced AI, holding fixed which AI capabilities exist and the extent to which they have proliferated; together with
    \item Building \textit{resilience }to advanced AI, defined as the capacity to adapt.
\end{itemize}

Some authors have flagged the importance of something very similar to what we call “adaptation” using a different term, viz. “defense”. For example,~\citep{kapoor2024impact} suggest that “assuming that risks exist for the misuse … in question, misuse analyses should clarify how society (or specific entities or jurisdictions) defends against these risks. … [N]ew defenses can be implemented or existing defenses can be modified to address the increase in overall risk.” In one respect, Kapoor et al.’s scope is narrower than ours: they focus on defence against risks arising specifically from \textit{misuse }of an advanced AI system by bad actors, whereas we urge consideration also of unintended harms via systemic effects and from loss of control. However, the underlying concept of “defence” appears to be similar or identical to our concept of “adaptation”.

Similarly,~\citep{krier2024models} discusses using frontier models to “improve societal defenses” against attacks that could be facilitated by advanced AI in the hands of bad actors; in this connection, he mentions the possibility of adaptive initiatives, including enhanced cybersecurity and closing legal loopholes. Krier’s focus is still narrower than that of Kapoor et al. since he focuses specifically on \textit{using frontier models to }improve defences, but again the underlying concept of improving defences seems similar to our concept of adaptation (or perhaps, in Krier’s case, specifically to the “defence” component thereof).

In addition, informally we are aware of several groups considering various nearby concerns under the heading of “AI resilience”, though we have not yet seen any corresponding sustained discussion in print.

\section{Adverse Impacts Statement}
\label{id:h.cym3cc7mzjbt}

This paper aims to positively affect how society responds to the opportunities and risks presented by advanced AI, via an increased and more well-targeted focus on adaptation.

Our primary concern is that this work might be misconstrued as a call to de-emphasise capability-modifying interventions. While we have argued that capability-modifying approaches have limitations in the long run, we believe they continue to be a crucially important component of risk management as frontier AI capabilities continue to be developed and deployed. Our argument is that we should \textit{also} invest seriously in adaptation measures.

In particular, like capability-modifying interventions, adaptation interventions are not failsafe guarantees of zero harm. Successful implementation of adaptation measures should not give free reign to AI developers to make risky deployment decisions.

\begin{ack}

We would like to thank the following for productive conversation and comments on previous drafts of the paper: Shahar Avin, Mauricio Baker, Matthew Bradbury, Ben Garfinkel, Josh Goldstein, Lujain Ibrahim, Nitarshan Rajkumar, Ben Robinson, Girish Sastry, Toby Shevlane, Merlin Stein and Jess Whittlestone, as well as participants in the Centre for the Governance of AI 2024 Winter Fellowship. 

\end{ack}

\bibliography{references}

\end{document}